\documentclass[11pt]{article}

\usepackage[top=1.5in, bottom=1.5in, left=1.5in, right=1.5in]{geometry} 

\begin{document}
\title{Comparing cost and performance of \\replication and erasure coding}
\author{John D. Cook\footnote{Singular Value Consulting. Houston, TX} \\
Robert Primmer\footnote{Hitachi Data Systems. Waltham, MA}, Ab de Kwant\footnote{Hitachi Data Systems. Herikerbergweg, Netherlands}}
\date{July 19, 2013}
\maketitle
\parindent=0in
\parskip=0.1in

\begin{abstract}
Data storage systems are more reliable than their individual components. In order to build highly reliable systems out of less reliable parts, systems introduce redundancy. In replicated systems, objects are simply copied several times with each copy residing on a different physical device. While such an approach is simple and direct, more elaborate approaches such as erasure coding can achieve equivalent levels of data protection while using less redundancy. This report examines the trade-offs in cost and performance between replicated and erasure encoded storage systems.
\end{abstract}

\section{Introduction}

To protect data from device or system failures storage systems employ mechanisms for data protection so that data can be provided to clients in the face of failure. There are a variety of means for providing data protection (DP), in this paper we will focus on two: \emph{replication} and \emph{erasure coding}. 

Providing data protection necessarily results in the consumption of additional storage. The additional storage needed to implement a given DP scheme is referred to as \emph{storage overhead}. A DP mechanism that uses less overhead is said to be more \emph{storage efficient}, meaning it requires less storage overhead to provide equivalent data protection. 

Replication is a process where a whole object is replicated some number of times, thus providing protection if a copy of an object is lost or unavailable. In the case of replicating a whole object, the overhead would be 100\% for a single replica, 150\% for two replicas and so forth. Erasure coding is a process where data protection is provided by slicing an individual object in such a way that data protection can be achieved with greater storage efficiency --- that is, some value less that 100\%.  

Most papers on erasure coding focus on the relative storage efficiency of erasure coding (EC) versus replication. However, these papers often ignore details of EC implementation that have an impact on performance, nor do they address the issue of data availability. 

Replication is a special case of EC, using only one data disk. The parity disks in this case are necessarily replicas. However, EC typically refers to other configurations, varying the number of data disks and parity disks to achieve the same amount of protection against data loss with fewer disks or to achieve greater protection against data loss with the same number of disks. However, an EC system may not provide as much protection against data unavailability relative to a replicated system with the same protection against data loss. An erasure coded system may also have more latency than a comparable replicated system.

In this paper we demonstrate that EC is advantageous for ``cold'' data --- data which is unlikely to be accessed or for which read performance is not an issue, and that replication is superior for ``hot'' data --- content which is accessed frequently or for which performance is important. This paper gives guidance for choosing between replication and erasure coding in the large space between ``hot'' and ``cold'' data. In particular, the more emphasis one places on availability and read performance, the greater the advantage of replication; the more emphasis one places on storage efficiency, the greater the advantage of erasure coding. 

\section{Data availability}

Enterprise-class data storage systems are designed to protect against data loss (DL) and data unavailability (DU). Data loss refers to permanent loss, data that cannot be retrieved by any means. Data unavailability refers to data that is temporarily unavailable but that can be retrieved after some undetermined delay. Said another way, DL and DU refer to data being permanently and temporarily inaccessible.

Cold data is often stored on tape due to the low cost of tape as a storage medium. The downside of storing data on tape is that it is generally less readily available compared to data stored on disk. Therefore, for cases where availability is important the increased expense of disk storage is justified. 

For data to be \emph{highly available} in a multi-data center environment means that the data must be distributed amongst the various data centers (DCs) in such a way that all data can still be read in the event that any one DC is unavailable. 

Data availability calculations are sometimes incorrectly conflated with data loss calculations. While the two can be identical for a single site, they diverge when we talk about multisite. For example, assume we have two DCs and want our data to be highly available, what is the \emph{minimum} amount of overhead required to protect against DU? The answer is 100\%, and this answer is the same irrespective of the DP scheme employed to protect against DL.

This answer can feel counterintuitive at first blush as it may seem that each DC need only house 50\% of the total data. However when we consider that for clients to be able to read the data when one of the data centers is unavailable requires that \emph{all} of the data be available from the only data center that's reachable, the answer makes sense. The table below illustrates how data must be spread across DCs to provide full data availability (i.e. full data access in the event of the failure of any one DC), independent of the data protection scheme employed. 
 
\begin{tabular}{l*{2}{c}l}
DCs & Minimum overhead for N+1 \\
\hline
2 & 100\% (100/100) \\
3 & 50\% (50/50/50) \\
4 & 33\% (33/33/33/33) \\
5 & 25\% (25/25/25/25/25) \\
\end{tabular}

\section{Data protection}

Data loss can occur when either a disk has died or when a disk is still working, but the blocks that hold the data being read have gone bad. For our purposes we'll just consider either case to have the same effect on our system. 

Let $p_L$ be the probability that a single disk is dead, permanently unreadable. Once a disk has been replaced, its function has been restored and we could think of the original disk coming back to life. Data loss does not necessarily happen when disks die but when too many disks are in a dead state simultaneously.

An estimate of $p_L$ depends on the mean time between failures as well as the time required to replace the disk. For example, suppose a disk performs reliably on average for 1000 days, roughly three years. If a hot spare is available to replace the disk upon failure and it takes one day to format the disk and copy data to it, $p_L = 0.001$. If there isn't a hot spare and it takes a couple days to order and install a replacement disk and a day to fill it, $p_L = 0.003$. 

Similarly let $p_U$ be the probability that a disk is unavailable. A disk may be unavailable yet still alive. For example, a disk may be in perfect order but temporarily disconnected from the network. Because disks can be temporarily unavailable without failing, say due to a network outage, $p_L$ is always smaller than $p_U$.

\subsection{Replication}

In a replicated system with $k$ replicas of each object, the probability of data loss is $p_L^k$, \emph{assuming disk failures are independent}. (We will discuss this independence assumption later.) Given a tolerable probability of data loss $\varepsilon$, we can solve for the number of disks $k$ needed keep the probability of loss below this level:

\begin{equation}
 k = \frac{\log \varepsilon}{\log p_L} \label{kreplication}
\end{equation}

The same calculation applies to data unavailability if we replace $p_L$ with $p_U$, again assuming independence. Some causes of unavailability may have independent probabilities, but as we discuss in Section \ref{DC}, disks located within a data center risk simultaneous unavailability due to the data center going offline temporarily.

\subsection{Erasure coding}

In an $m+n$ erasure encoded system, each object is divided into $m$ equal-sized fragments. In addition $n$ parity fragments are created, each the size of one of the data fragments. The data can be read if any $m$ out of the $m+n$ fragments are available. Stated negatively, data will be lost if more than $n$ fragments are simultaneously lost. In this section we will focus on data loss. Similar considerations apply for data unavailability, though there is an important distinction between data loss and data unavailability that matters more for than for replication. More on this below.

Since we need $m$ out of $m+n$ disks to reconstruct an object, the probability of data loss is the probability of more than $n$ disks being dead simultaneously:

\begin{equation}
\sum_{i=n+1}^{m+n} {m+n \choose i} p_L^i (1-p_L)^{m+n-i} \label{ecfailure1}
\end{equation}

Define $k_c = (m+n)/m$. This is the redundancy factor for $m+n$ erasure coding, analogous to $k$ for replication. For a fixed redundancy level, an erasure encoded system has at least the same level of reliability as the corresponding replicated system. (In general, an erasure encoded system will have greater reliability, but one could consider replication as a special case of erasure coding with $m = 1$ in which case of course the systems would have equal reliability.) Stated another way, for a fixed level of reliability, an erasure encoded system requires less disk space or can store more data with the same amount of disk space.

Given a number of data fragments $m$ and an acceptable probability of unavailability $\varepsilon$, you can solve for the smallest value of $n$ such that
\begin{equation}
\sum_{i=n+1}^{m+n} {m+n \choose i} p_L^i (1-p_L)^{m+n-i} < \varepsilon \label{ecfailure2}
\end{equation}
See Appendix $A$ for a Python script to solve for $n$. 

For example, suppose you would like to build a system with probability of data recovery 0.999999 (six nines) using disks that have probability 0.995 of being alive. Triple replication would have a probability of DL equal to $0.005^3 = 1.25 \times 10^{-7}$. Suppose you want to use erasure coding with 8 data disks. The script in the appendix shows that an $8+n$ system would require $n=3$ to keep the probability of DL below $10^{-6}$. In fact, an $8+3$ system has a probability of DL $1.99 \times 10^{-7}$. A 1 GB video stored in the triple replicated system would require 3 GB of storage. In an 8+3, the same object would be stored in 8 data fragments and 3 parity fragments, each 125 MB in size, for a total of 1.375 GB. In short, the erasure coded system would use about half as much disk space and offer the same level of data protection.

(The number of parity disks $n$ is best calculated numerically as mentioned above, though there are (incorrect) analytical solutions in the literature. Rodrigues and Liskov \cite{rl} give an analytical solution for (\ref{ecfailure2}) by using a normal (Gaussian) approximation to the binomial distribution Bin($m+n$, $p$). However, the parameters $m+n$ and $p$ are typically outside the range for the normal approximation to be reliable. The estimate for $k_c$ given in that paper has a large error. In the example above where we concluded 3 parity disks were needed, the approximation of Rodrigues and Liskov says 1.054 disks are needed. Even if you round 1.054 up to 2, this is still short of the 3 disks necessary.)

\subsubsection{Choosing the number of data disks}

Given a value of $m$ and the individual disk reliability, the discussion above describes how to choose $n$ to achieve the desired level of protection. But how do you choose $m$?

Increasing $n$ while holding $m$ fixed increases reliability. Increasing $m$ while holding $n$ fixed \emph{decreases} reliability. But in a sense, we gain more reliability by increasing $n$ than we lose by increasing $m$. We can increase reliability by increasing $m$ and $n$ proportionately, this keeping the redundancy factor $k_c$ constant. See Appendix $B$ for a precise statement and proof.

For example, a $4+2$ system will be more reliable than a $2+1$ system even though both have the same redundancy $k_c = 1.5$. So why not make $m$ and $n$ larger and larger, obtaining more and more reliability for free? Of course the increase in $m$ and $n$ is not free, though the cost is subtle.

Larger values of $m$ and $n$ do result in greater data protection for a fixed level of disk reliability. However, while larger values of $m$ and $n$ reduce the chance of complete failure (irrecoverable data loss), they \emph{increase} the chance of at least one recoverable disk failure. As we will describe below, this potentially increases latency and reconstruction costs.

Increasing $m$ and $n$ also increases the total number of data fragments $m+n$ to manage. In practice, erasure encoded systems use values of $m$ on the order of 10, not on the order of 100.  For example, you may see $6+3$ systems or $12+4$ systems, but not $100+50$ systems. One reason is that you can obtain high levels of data protection without resorting to large values of $m$. Another is that erasure encoded systems typically have on the order of millions of object fragments. A rough calculation shows why this is so. There must be a database to keep track of each fragment. If this database is kept in memory, it has to be on the order of gigabytes. If each record in the database is on the order of a kilobyte, the table can contain on the order of a million rows, and so the number of fragments is kept on the order of millions in order to fit the corresponding database in memory.

Aside from the memory required to keep an inventory of data fragments, there is also the time required to find and assemble the fragments. The required time depends greatly on how EC is implemented, but is always some amount of time, and hence an overhead associated with EC that is not required with replication. This explains why an EC system can be slower than a replicated system, even when all fragments are in the same data center.

Finally, we note that the more data fragments there are to manage, the more work required to rebuild the fragment inventory database when failures occur.

\subsection{Allocating disks to data centers} \label{DC}

The previous sections have not considered how disks are allocated to data centers. This allocation impacts EC more than replication, and DU more than DL.

It is far more likely that an entire data center would be unavailable than that an entire data center would be destroyed. Data centers are occasionally inaccessible due to network outages. This causes all disks inside to be simultaneously unavailable. However, it is very unlikely that all disks in a data center would fail simultaneously, barring a natural disaster or act of war. This means that DU is more correlated than DL.

In replicated systems, it is common for one replica of each object to reside in each data center. If we assume network failures to data centers are independent, then the same probability calculations apply to data loss and data unavailability. If there are $d$ data centers, the probabilities of an object being lost or unavailable are $p_L^d$ and $p_U^d$ respectively.

However, in EC systems, the number of fragments per object is typically larger than the number of data centers. A company often has two or three data centers; more than four would be very unusual, especially for a mid-sized company. And yet EC systems using a scheme such as 8+4 are not uncommon. With fewer than 12 data centers, some of these fragments would have to be co-located.

If every fragment in an EC system were stored in a separate data center, the unavailability calculations would be analogous to the data loss calculations, as they are for replicated systems. But because data fragments are inevitably co-located, these fragments have correlated probabilities of being unavailable and so the unavailability probability for the system goes up. 

\subsection{Probability assumptions}

Reliability calculations, whether for replicated systems or erasure encoded systems, depend critically on the assumption of independence. Disk failures could be correlated for any number of reasons -- disks coming from the same manufacturing lot, disks operating in the same physical environment, etc. -- and so reliability estimates tend to be inflated \cite{goog}. Researchers from Carnegie Mellon have shown that, for example, that disk failures are correlated in time \cite{cmu}. That is, if you've had more failures than usual this week, you're likely to have more failures than usual next week too, contrary to the assumption of independence. To be cautious one should look for ways to improve independence and be skeptical of extremely small probabilities calculated based on independence assumptions. 

The assumption of independence is more accurate for disks in separate data centers. And so for replicated systems with each replica in a separate data center, independence is a reasonable assumption. But for EC systems with multiple fragments in each data center, the assumption of independence is less justified for data loss, and unjustified for data unavailability.

Also keep in mind that when the probability of any cause of failure is sufficiently small, other risks come to dominate. One could easily build an erasure coded system with theoretical probability of failure less than $10^{-12}$, a one in a trillion chance. But such probabilities are meaningless because other risks are much larger. Presumably the chance of an earthquake disabling a data center, for example, is larger than $10^{-12}$.

\section{Efficiency considerations}

In a replicated system, the operating system directly locates objects. In the event of a failure, requests are redirected to another server, but in the usual case objects are accessed directly. 

With erasure coding, fragments of objects must be cataloged. Since the bits necessary to reconstruct an object exist on multiple disks, a system must keep track of where each of these fragments are located. When an object is requested from an $m+n$ system, a server looks up the location of at least $m$ fragments. (The details of this process are implementation dependent. A server could, for example, randomly choose $m$ fragments or try to determine the $m$ closest fragments.) If the requests succeed, these $m$ fragments are transferred to a server to be re-assembled into the requested object. Since assembly cannot begin until the last fragment is available, the process would be slowed down if one of the fragments were coming from a location further away than the others.

\section{Costs of disk failures}

For a fixed level of reliability, erasure coding requires less disk space than replication. If storage cost were the only consideration, erasure coding would have a clear advantage over replication. However, there are other factors to consider, and these factors depend on usage scenarios.

While erasure coding can lower the probability of a catastrophic failure, it increases the probability of a recoverable failure. The probability that one or more disks in a set will fail is approximately the probability of a particular disk failing times the number of disks.\footnote{The probability that \emph{exactly} one out of a set of $m$ disks will fail is $mp$ if $p$ is the probability of an individual failure. But the probability of one \emph{or more} failures is $1-(1-p)^m$ which is approximately $mp$ if $p$ is small.} An example given above compared triple replication to $8+3$ erasure coding. Both systems had roughly the same probability of data loss. Replication used 3 disks per object while erasure coding used 11, and so the erasure coded system is nearly 4 times as likely to experience a single, recoverable disk failure.

If a disk failure is recoverable, what is its cost? The cost is not in maintenance. The cost of replacing hard disks is proportional to the total number of disks in use, whether those disks contain replicated objects or erasure encoded object fragments. Erasure encoding does not increase the number of disk failures. By reducing the number of disks needed, it can \emph{reduce} the number of failures. However, erasure encoding increases the probability that an object request is effected by a disk failure, and this may increase latency.

\subsection{Latency}

If all replicas and all erasure encoded fragments are in the same data center, latency is not as much of an issue as when replicas and encoded fragments are geographically separated. 

In a replicated system, requests could be routed to the nearest available replica (nearest in terms of latency, which usually means physically nearest though not necessarily). If the nearest replica is not available, the request would fail over to the next nearest replica and so on until a replica is available or the request fails. In an $m+n$ erasure encoded system, an object request could be filled by reconstructing the object from the $m$ nearest fragments. Since object requests more often encounter (recoverable) disk failures in an erasure encoded systems, they will more often involve an increase in latency.

Suppose a replicated system maintains $k$ copies of an object, each in a different data center, and that requesting data from these centers has latency $L_1 < L_2 < \ldots < L_k$ for a given user. Suppose each replica has a probability $p$ of being unavailable. With probability $1-p$ the latency in the object request will be $L_1$. With probability $p(1-p)$ the latency will be $L_2$. The expected latency will be

\[ \sum_{i=1}^k p^{i-1}(1-p) L_i \]

If $p$ is fairly small, the terms involving higher power of $p$ will be negligible and the sum above will be approximately

\[ (1-p)L_1 + p L_2\]

dropping terms that involve $p^2$ and higher powers of $p$.

For example, if there is a probability 0.001 that an object is available at the nearest data center, and the second data center has 100 times greater latency, the expected latency is 10\% greater (i.e. $p L_2/L_1 = 0.001*100 = 0.1$) than if both replicas were located in the nearest data center.

In an erasure encoded system, fragments could be geographically distributed in many ways. If latency is a concern, an $m+n$ system would store at least $m$ fragments in a single location so that only local reads would be necessary under normal circumstances. If $m$ fragments are stored in one location, the probability of one local disk failure is $mp$. This means the probability of a single local failure, and the necessity of transferring data from a more distant data center, is $m$ times greater for an erasure coded system compared to a replicated system. The expected latency increases from approximately $(1-p)L_1 + pL_2$ in a replicated system to approximately $(1-p)L_1 + mpL_2$ in an erasure encoded system. 

To illustrate this, suppose in the example above that we pick our units so that the latency of the nearer server is $L_1 = 1$. If $m = 8$, the expected latency would be 
1.099 using replication and
1.799 using erasure coding. 

For active data, objects that are accessed frequently, latency is a major concern and the latency advantage of replication should be considered. For inactive data, objects are archived and seldom accessed, latency may be less of a concern.

\subsection{Reconstruction}

When an object request fails, it must be determined exactly what has failed. This is a much simpler task in a replicated system than in an EC system.

Once the failed disk has been identified, in either a replicated or an erasure coded system, its data must be copied onto a replacement disk. The cost of rebuilding the disk depends on where the data are coming from, whether locally within a data center or remotely from another center. 

With replication, the content of a failed disk is simply copied, either from within a data center or from a remote data center, depending on how replicas are distributed.

With $m+n$ erasure coding, the content of $m$ disks must be brought to one location. Huang \emph{et al} \cite{lrc} give the example of $6+3$ encoding. If three fragments are stored in each of three data centers and one disk fails, the content of four disks must be transferred from a remote data center to the site of the failed disk in order to have enough data for reconstruction. 

The authors propose \emph{local reconstruction codes} \cite{lrc} as a variation on erasure codes to mitigate this problem. With this approach, two data centers would each contain three data disks and a local parity disk. A third data center would contain two global parity disks computed from all six data disks. They call this approach 6+2+2. Any single failure could be repaired locally. Since single failures are most likely, this reduces the average reconstruction time. The $6+2+2$ scheme offers a level of data protection intermediate between $6+3$ and $6+4$ Reed-Solomon codes. The $6+2+2$ system can recover from any combination of three-disk failures, and from 86\% of four-disk failures.

The cost of reconstructing a disk is lowest with replication, highest with traditional Reed-Solomon erasure coding, and intermediate with local reconstruction codes.

The time necessary for reconstruction feeds back into the data protection calculations. If failed disks can be reconstructed faster, availability improves, increasing the availability of the system or possibly enabling the system to use fewer disks.

\section{Trade-offs}

In a system with a large volume of active data -- objects accessed frequently -- and in which it is important to minimize latency, replication has the advantage. A content delivery network (CDN), for example, would use replication over erasure coding. 

On the other hand, a system with mostly inactive data -- archival objects accessed rarely -- and primarily concerned with storage costs, would be better off using erasure coding. Such a system might also want to use a moderately large value of $m$ in its $m+n$ encoding.

A hybrid approach, one used by Microsoft Asure \cite{lrc}, is to use both replication and erasure coding. Objects are replicated on entering the system. Later they are erasure encoded and the replicas are deleted. This way active data is served by replication and inactive data is served by erasure coding. 

In the Azure implementation, the cost of encoding fragments is not a significant concern. Because objects are replicated immediately, encoding can be done out of band. A system that encodes objects as they enter might be more concerned about encoding costs.

One could also use a hybrid approach of using $k$ replications of $m+n$ erasure encoded fragments. In such a case, $k$ and $n$ might be very small. For example, $n$ might be 1 or 2, and $k=2$ or 3 might be the number of data centers. 

\section{Recommendations}

In summary, we recommend beginning by estimating the reliability of individual components and the desired reliability of the overall system. Then one could use equation (\ref{kreplication}) to determine the necessary number replicas $k$ or use equation (\ref{ecfailure2}) to determine the number of fragments in an $m+n$ erasure coded system. 

Replicated systems are simpler to implement and can have lower latency and reconstruction costs. Erasure encoded systems require less disk space and hence have lower storage costs. A system primarily used for serving active data as quickly as possible would probably use replication. A system primarily used for archiving would benefit from erasure coding. A system with mixed use might choose a compromise, such as erasure coding with a small number of data fragments per object, depending on the relative importance of latency, reconstruction costs, and storage/maintenance costs. Or such a system could use both replication and erasure coding, moving objects from the former to the latter as they are accessed less frequently.

\section{Appendix $A$: Software} 

\begin{verbatim}
from math import sqrt, factorial

def binomial(a, b):
    return factorial(a + b)/(factorial(a)*factorial(b))

def prob_fail(p, m, n):
    """Probability of more than n failures in an m+n system 
    when each component has probability of failure p."""
    sum = 0
    for i in range(n+1, m+n+1):
        sum += binomial(m+n, i)*p**i*(1-p)**(m+n-i)
        return sum

def parity_disks_needed(epsilon, p, m):
    """
    Solve for number of parity disks to achieve specified reliability.
    epsilon = acceptable probability of data loss
    p = probability of single disk failure
    m = number of data disks
    """

    n = 0
    while True:
        n += 1
        if prob_fail(p, m, n) < epsilon:
            return n  
\end{verbatim}

\section{Appendix $B$: Increasing $m+n$ increases reliability} 

Let $m$ and $n$ be fixed positive integers. Let $p$ be the probability of a single disk being dead at any time. Assume $p < n/(m+n)$.

We start with an $m+n$ EC system and increase $m$ and $n$ proportionately, resulting in a $km + kn$ system. As we increase $k$, the probability of data loss, i.e the probability of more than $kn$ failures, goes to zero as $k$ increases.

To see why this is so, let $X$ be a Binomial($km + kn$, $p$) random variable and let $Y$ be a normal random variable with mean $k(m+n)p$ and variance $k(m+n)p(1-p)$. As $k$ increases, the probability distribution of $X$ converges to the probability distribution of $Y$. 

For any positive $k$, 
\[ \mathrm{Prob}(X > kn) \approx \mathrm{Prob}(Y > kn) \]
and the error in the approximation decreases with $k$, becoming exact in the limit. Let $Z$ be a standard normal random variable. Then $Y$ has the same distribution as 
\[\sqrt{k(m+n)p(1-p)} Z  + k(m+n)p\]
and so we have the following.

\begin{eqnarray*}
\mathrm{Prob}(X > kn) &\approx& \mathrm{Prob}(Y > kn) \\
&=& \mathrm{Prob}(\sqrt{k(m+n)p(1-p)} Z + k(m+n)p > kn) \\
&=& \mathrm{Prob}\left(Z > \sqrt{k} \frac{(1-p)n - mp}{(m+n)p(1-p)}\right) 
\end{eqnarray*}

The assumption $p < n/(m+n)$ implies that $(1-p)n - mp > 0$ and so 
\[ \lim_{k\to\infty} \mathrm{Prob}\left(Z > \sqrt{k} \frac{(1-p)n - mp}{(m+n)p(1-p)}\right) = 0.\]

\end{document}